\documentclass[aps,prl,twocolumn]{revtex4}
\usepackage[usenames]{color}
\newcommand{\comment}[1]{}
\usepackage{graphicx}
\usepackage{epsfig}
\newcommand{\bra}[1]{\langle {#1} |}
\newcommand{\ket}[1]{| {#1} \rangle}
\newcommand{\expect}[1]{\langle {#1} \rangle}
\newcommand{\ketn}[1]{ {#1} \rangle}

\begin{document}

\bibliographystyle{revtex}
\title{Classifying Novel Phases of Spinor Atoms}
\author{Ryan Barnett$^{1,2}$, Ari Turner$^1$, and Eugene Demler$^1$}
\date{\today}
\affiliation{$^1$ Department of Physics,
  Harvard University, Cambridge, Massachusetts 02138, USA}
\affiliation{$^2$ Department of Physics, California Institute of Technology, MC 114-36, Pasadena, California 91125, USA}
\date{\today}

\begin{abstract}
We consider many-body states of bosonic spinor atoms which, at the
mean-field level, can be characterized by a single-particle wave
function for the BEC and Mott insulating states. 
We describe and apply a  
classification scheme
that makes explicit spin symmetries of such states and enables one to
naturally 
analyze their collective modes and topological excitations. 
Quite generally, the method allows classification of a spin $F$ system
as a polyhedron with $2F$ vertices.
We apply the method to the
many-body states of bosons with spins two and three. For
spin-two atoms we find the ferromagnetic state, a continuum of 
nematic states, and a state having the symmetry of the point
group of the regular tetrahedron. For spin-three atoms we obtain similar
ferromagnetic and nematic phases as well as 
states having symmetries of various types of
polyhedra with six vertices: the hexagon, the pyramid with pentagonal base,
the prism, and the octahedron.
\end{abstract}

\maketitle

% Begin introduction
Ultracold atoms in either a single optical trap or in an optical lattice
provide clean realizations of unique systems of spins
which were previously studied
only as toy mathematical problems 
(for a recent review, see \cite{lewenstein06}).
Depending on which hyperfine state is populated, alkali atoms
can  have spin one or two.
Different phases of spin-one alkali
bosons have been experimentally realized
\cite{stenger98,schmaljohann04, chang04, chang05}
and considered theoretically
\cite{ho98,ohmi98,zhou04,imambekov03,garcia-ripoll04,zhang04,bernier05}.
Spin-two bosons have also been experimentally probed
\cite{schmaljohann04, chang04, widera05,gerbier06}
and theoretically studied for the case
of a single optical trap \cite{ciobanu00,ueda02}
as well as, very recently, an optical lattice  \cite{zawitkowski06,zhou06}.
Finally, the Stuttgart group has recently succeeded in
obtaining a Bose-Einstein condensation of $^{52}$Cr atoms
\cite{griesmaier05} which are spin-three bosons.
This was followed by theoretical work
\cite{santos06,diener06} showing the possible types of phases
that can be realized for such a spin-three system.

Classification schemes of single-particle
states with non-zero spins are needed to describe both superfluid
condensates and Mott insulating states of  spinor bosonic atoms.
However, such classification
becomes increasingly more difficult for larger spins.
For instance,  classifying the state of a spin-half
particle is straightforward;
only knowledge of the expectation value
of the spin operator $\expect{\bf F}$ is needed.
On the other hand, for a spin-one particle,
knowledge of the expectation value of the nematic tensor
familiar from the classical theory of liquid crystals \cite{degennes95}
\begin{equation}
\label{Eq:nematic}
Q_{ab}=\frac{1}{2}(F_a F_b + F_b F_a) - \frac{1}{3} \delta_{ab} F^2
\end{equation}
in addition to $\expect{\bf F}$ is required.
Proceeding along these lines, one finds that for larger
spin such a classification scheme becomes
quite cumbersome since one needs  to consider
order parameters that involve higher-order products of spin operators,
and a
physical interpretation is not immediate.
% End introduction

% Begin summary
In this Letter, we  present an alternative classification scheme which will work well for large spin.  This scheme allows the symmetries of a general spin $F$ particle to be represented by a polyhedron with $2F$ vertices. To illustrate the method we use it to discuss spin-two bosons which are naturally realized as a hyperfine state of alkali atoms. The previously discussed superfluid ferromagnetic, polar, and cyclic condensates \cite{ciobanu00,ueda02} are shown to have the transformation properties of a \emph{ferromagnet}, a \emph{nematic}  (either uniaxial or biaxial),  and a \emph{tetrahedron}.  In addition, we show that the nematic phase has an additional degeneracy at the mean-field level. We then show that states having similar transformation properties will be obtained in the Mott insulating state. Next, we show that the classification scheme immediately gives the number of Goldstone modes and allows one to classify topological excitations. In fact, in the biaxial nematic phase and the tetrahedratic phase, the topological excitations combine according to nonabelian groups \cite{mermin79}.  Finally, due to recent experimental interest, we will also discuss classification of spin-three bosons.
% End summary

% Begin general method
We will now outline our general classification scheme
for a single spin.
The key ingredient of our classification scheme is identifying states of
spin $F$ particles with $2F$ points on the unit sphere \cite{bacry74}.
Consider a particle of
spin $F$ in the state given by
$\ket{\psi}=\sum_{\alpha=-F}^{F} A_{\alpha} \ket{\alpha}$
where $F_z\ket{\alpha}=\alpha \ket{\alpha}$ and $A_\alpha$
are a set of normalized complex coefficients.
To gain a physical understanding of this state,
the idea is to find the set of ``maximally polarized'' states which are
orthogonal to $\ket{\psi}$.  The maximally polarized state
$\ket{\zeta}$ pointing in the $\hat{\bf n}=(\theta,\phi)$ direction
is determined by the equation
$
{\bf F} \cdot \hat{\bf n} \ket{\zeta}= F \ket{\zeta}.
$
A convenient (not normalized) representation of this state
(which is related to the Schwinger boson representation) is
\begin{equation}
\label{Eq:pol}
\ket{\zeta}=\sum_{\alpha=0}^{2F} \sqrt{2F \choose \alpha}
\zeta^{\alpha} \ket{F-\alpha}
\end{equation}
where
$
\label{Eq:stereographic}
\zeta=e^{i\phi} \tan(\theta/2)
$
is the stereographic
mapping of the unit sphere to the complex plane.
We define  the characteristic polynomial for $\ket{\psi}$
in $\zeta$ to be
\begin{equation}
\label{Eq:characteristic}
f_{\psi}(\zeta)\equiv \bra{\psi}\ketn{\zeta}=
\sum_{\alpha=0}^{2F} \sqrt{2F \choose \alpha} A^{*}_{F-\alpha}
\zeta^{\alpha}.
\end{equation}
The values of the $2F$ complex roots of $f(\zeta)$,
$\{ \zeta_i \}\equiv \{\zeta : f_{\psi}(\zeta)=0 \}$,
(which are in one-to-one correspondence with a
set of points on the unit sphere $\{ (\theta_i,\phi_i) \}$)
determine the coefficients
$A_\alpha$ and therefore $\ket{\psi}$ up to
normalization and an overall phase factor.  \emph{Most importantly,
the symmetries of $\ket{\psi}$ correspond to the operations
under which the set of points on the unit sphere
$\{ (\theta_i,\phi_i) \}$ are invariant.}  A similar method
has been developed in the $19th$ century
mathematics community to solve quintic
polynomials in terms of rotation of regular icosahedra
\cite{klein03}.
% End general method

% Begin single trap BEC part
We illustrate our general approach by considering 
a specific problem of spin-two bosons, first
concentrating on the BEC state in a single optical trap and then extending 
the discussion
to the insulating Mott state in an optical lattice.
Spin two bosons interact with the contact potential
\begin{eqnarray}
V_{\rm int}({\bf x_1}-{\bf x_2})=\delta({\bf x_1}-{\bf x_2})
(g_0 {\cal P}_0 + g_2 {\cal P}_2 + g_4 {\cal P}_4)
\label{Vcontact}
\end{eqnarray}
where ${\cal P}_F$ projects into the state with
total spin $F$ and $g_F=2\pi a_F/m$ where $a_F$ is the scattering
length corresponding to spin $F$.
Note the requirement that the wave functions
be symmetric under interchange of particles prohibits odd spin
projection operators.
In the superfluid phase when all atoms occupy the same orbital state,  potential (\ref{Vcontact})
leads to the interaction Hamiltonian
\cite{ciobanu00, ueda02,zawitkowski06} of
\begin{equation}
\label{Eq:Hint}
{\cal H}_{{\rm int}} = \frac{1}{2} U_0 n(n-1) +
\frac{1}{2} U_1 {\cal P}_{0}
+ \frac{1}{2} U_2 (F^2- 6n)
\end{equation}
where
$U_0=\frac{\gamma}{7}(4 a_2 + 3 a_4)$,
$U_1=\frac{\gamma}{7}(7 a_0 - 10 a_2 + 3 a_4)$, and
$U_2=\frac{\gamma}{7}(a_4 - a_2)$ ($\gamma$ is a positive constant).
In Eq.~(\ref{Eq:Hint})
$n=a_{\alpha}^\dagger a_{\alpha}$ (here and after sums over repeated
indices are implied) where $a_{\alpha}^\dagger$ creates a spin-two
boson in an eigenstate of $F_z$ having eigenvalue $\alpha$.
Finally, $F^2={\bf F}\cdot {\bf F}$ where
${\bf F}=a^{\dagger}_{\alpha} {\bf T}_{\alpha \beta} a_{\beta}$ and
${\bf T}_{\alpha \beta}$ are the spin-two matrices.

\begin{figure}
\includegraphics[width=3.5in]{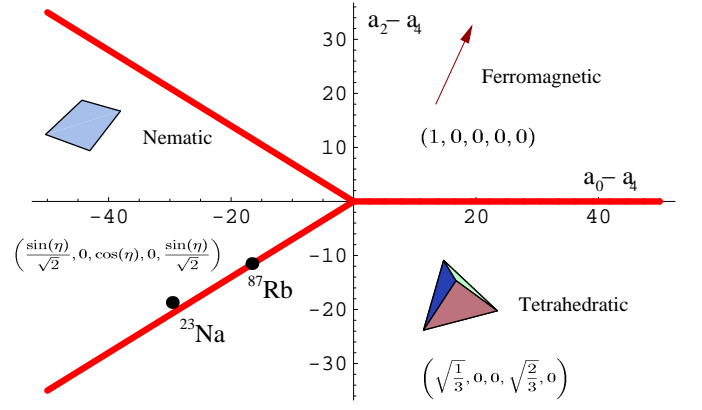}
\caption{Phase diagram for spin-two bosons in a single optical
trap. The axes are in units of $a_B$.
For the Mott insulating state with one bosons per site,
the phase diagram will be the same
with horizontal and vertical axes corresponding to $\epsilon_0-\epsilon_4$ and
$\epsilon_2-\epsilon_4$ respectively (see Eq. \ref{Eq:Heff}).}
\label{Fig:phased1}
\end{figure}

The mean-field description for a BEC of $N$ 
bosons is $\ket{\psi_{\rm SF}}=\frac{1}{\sqrt{N!}} \left( A_{\alpha} a_{\alpha}^{\dagger}\right)^N \ket{0}$,
where the normalized complex coefficients $A_{\alpha}$ are the variational
parameters.  Since the kinetic energy of the system will
not depend on the spin degrees of freedom, it is sufficient
to minimize $\bra{\psi_{\rm SF}} {\cal H}_{\rm int} \ket{\psi_{\rm SF}}$
over the variational parameters.
To leading order in $N$, we find with this wave function
\begin{eqnarray}
\expect{{\cal H}_{\rm int}}/N^2&=&
\frac{1}{10} U_1 |2A_2 A_{-2} - 2 A_{1} A_{-1} +A_0 A_0|^2
\nonumber
\\
&+& \frac{1}{2} U_0
+ \frac{1}{2} U_2 (A^{*}_\alpha {\bf T}_{\alpha \beta} A_{\beta})^2.
\end{eqnarray}
Carrying out the minimization with the constraint of normalization,
$A^{*}_{\alpha}A_{\alpha}=1$,
we find the
phase diagram given in Fig.~\ref{Fig:phased1} where the
boundaries are given by
$a_2-a_4=\pm \frac{7}{10}(a_0-a_4)$ and
$a_2=a_4$.
The phases correspond to the
coefficients $(A_{-2},A_{-1},A_{0},A_1,A_2)$ given by
\begin{eqnarray}
\label{Eq:variational}
\nonumber
&{\cal F:}& \; (1,0,0,0,0)
\\
&{\cal N:}& \; \left(\frac{\sin(\eta)}{\sqrt{2}},0,
\cos(\eta),0,\frac{\sin(\eta)}{\sqrt{2}}\right)
\\
\nonumber
&{\cal T:}& \;\left( \sqrt{\frac{1}{3}},0,0,
\sqrt{\frac{2}{3}} ,0\right)
\end{eqnarray}
which are uniquely determined
up to SO(3) spin rotations by the minimization.  Phase $\cal N$ has the
additional degeneracy parameter $\eta$ which does not correspond to a
symmetry of the hamiltonian.  One should note, however, that
this degeneracy only arises at the mean-field level and will be
removed when we include corrections to the mean-field wave function
\cite{turner06}.
On the other hand, in real experiments, the quadratic Zeeman
effect provides another way of removing this degeneracy which we will
discuss.
Using the scattering
lengths
taken from Ref. \cite{ciobanu00}
we also indicate where
$^{23}$Na and $^{87}$Rb will lie in the phase diagram.
The same phase boundaries were found by Ciobanu, Yip, and Ho \cite{ciobanu00}
for the superfluid phase.  However, we emphasize that there is a 
continuous degeneracy in the phase $\cal N$ at the mean-field level.

% End single trap BEC part

% Begin classification
Symmetries of the phases given by the wave functions in
Eq.~(\ref{Eq:variational}) is not immediate upon first glance.
We will now illustrate the utility of our method by
classifying the various phases.  For spin two, the maximally polarized
state as in Eq.~(\ref{Eq:pol}) is
\begin{equation}
\ket{\zeta}=\zeta^4 \ket{-2} + 2 \zeta^3 \ket{-1}
+ \sqrt{6} \zeta^2 \ket{0} + 2 \zeta \ket{1} + \ket{2}.
\end{equation}
Phase $\cal F$
is clearly the ferromagnetic state.
However, it will prove instructive to proceed with our systematic
classification scheme.
The characteristic polynomial
Eq.~(\ref{Eq:characteristic}) for this case is simply
$
f_{\psi_{\cal F}}(\zeta)=\zeta^4
$
which has the four-fold degenerate
root of $\zeta=0$ which corresponds to the north
pole of the unit sphere $(\theta,\phi)=(0,0)$.  The only symmetries
this state possesses are therefore the U(1) group of continuous rotations
about the $z$-axis.

We now move on to phase $\cal N$.  Before applying our classification
method, it is first useful to consider the order parameters constructed
in the conventional way.
First, we note that $\expect{\bf F}=0$.  Next the eigenvalues
of the nematic order
parameter Eq.~(\ref{Eq:nematic}) are
\begin{equation}
\label{Eq:nemEig}
{\rm Eig}(Q_{ab})=
\left(
\begin{array}{c}
-2\cos(2\eta)\\
 \cos(2\eta)+\sqrt{3}\sin(2\eta)\\
 \cos(2\eta)-\sqrt{3}\sin(2\eta)\\
\end{array}
\right).
\end{equation}
Now, applying our method the characteristic polynomial is
\begin{equation}
f_{\psi_{\cal N}}(\zeta)=
\sqrt{\frac{1}{2}} \sin(\eta) \zeta^4 +\sqrt{6}\cos(\eta)\zeta^2
+\sqrt{\frac{1}{2}} \sin(\eta).
\end{equation}
The roots of this equation all lie at the vertices of a rectangle.
For $0<\eta<\pi/3$, $\pi/3<\eta<2\pi/3$,
and $2\pi/3<\eta<\pi$ the rectangle will lie in the $zy$, $xy$,
and $xz$ planes respectively.   We find that $\eta=n\pi/3$ (defined ${\cal N}1$)  
corresponds
to two sets of double roots at opposite poles.  This
state will have the transformation properties of a uniaxial nematic.
That is, there is the symmetry of
continuous rotation
about the nematic axis.  There are also
the additional symmetries of discrete rotations by
$\pi$ about any axis in the plane perpendicular to the nematic axis.
On the other hand, when $\eta=(n+1/2)\pi/3$ (defined ${\cal N}2$)
the roots lie at the vertices of a
square which has no continuous symmetries (contrary to what one might be
led to believe from $Q_{ab}$).  The symmetries here are rotations
by $\pi/2$ about the nematic axis and rotations by $\pi$ about the
other two principal axes.
For all other values of $\eta$ the
nematic will be biaxial, which we refer to as phase ${\cal N}3$.
This case has the fewest symmetry operations, the rotations
by $\pi$ about the three principal axes.

A route to remove the mean-field
degeneracy associated with $\eta$ is to apply an external magnetic
field.  Since for typical experiments the time it takes a spin state
to relax is longer than the trap lifetime \cite{stenger98},
it is appropriate to think of $\expect{F_z}$ as a conserved quantity,
and the linear Zeeman term is unimportant.
Thus, the most important effect due to an external magnetic field
is due to the quadratic Zeeman term (proopertional to
$F_z^2$) which can either have a positive
or negative coefficient.  In the (most common) event that the coefficient 
is negative, the phase ${\cal N}2$ which transforms
as the square will be favored.  On the other hand, if the coefficient is
positive, the ${\cal N}1$ phase will be favored.

We now consider the remaining region in the phase diagram,
phase $\cal T$, which, as we will see,
is tetrahedratically ordered.
For this case, $Q_{ab}$ gives no useful information since
$\expect{Q_{ab}}=0$ as well as $\expect{\bf F}=0$.  Thus,
the utility of our classification method will be demonstrated
most clearly here.
The characteristic polynomial for this
situation is
$
f_{\psi_{\cal T}}(\zeta)=\sqrt{\frac{1}{3}} \zeta^4+2 \sqrt{\frac{2}{3}} \zeta.
$
The roots of this equation correspond to the points
$\{ (\theta_i,\phi_i) \}=\{ (0,0), (\bar{\theta}, \pi/3),
(\bar{\theta}, \pi), (\bar{\theta}, 5\pi/3) \}$
where $\bar{\theta}=2\tan^{-1}(\sqrt{2})$.  These points lie at the vertices of a regular tetrahedron
on the unit sphere.  Thus $\ket{\psi_{\cal T}}$ has the full
symmetry of the point
group of the tetrahedron.
% End classification

% Begin optical lattice part
We will now consider what happens when the spin-two bosons are in
an optical lattice in a Mott Insulating state (one per site).
Such a system will be described by the
hamiltonian
$
{\cal H}={\cal H}_{\rm kin} + {\cal H}_{\rm Hub}
$
where
$
{\cal H}_{\rm kin} = -J \sum_{\expect{ij}} (a^{\dagger}_{i \alpha} a_{j \alpha}
+ {\rm h.c.})
$
describes hopping between adjacent sites
and ${\cal H}_{\rm Hub}$ gives the on-site interaction
\begin{equation}
\label{Eq:Hub}
{\cal H}_{{\rm Hub},i} = \frac{1}{2} U_0 n_i(n_i-1) +
\frac{1}{2} U_1 {\cal P}_{0,i}
+ \frac{1}{2} U_2 (F_i^2- 6n_i).
\end{equation}
Note the similarity between the on-site interaction and
Eq.~(\ref{Eq:Hint}).
When the sites are completely decoupled, the ground state
energy will be independent of the spin state of any
given site leading to a macroscopic degeneracy which
will be removed by hopping.
Treating the hopping perturbatively, we arrive at the
effective hamiltonian
\begin{equation}
\label{Eq:Heff}
{\cal H}_{\rm eff}=
\sum_{\expect{ij}}
\left(\epsilon_0 {\cal P}_{0}(ij) +\epsilon_2 {\cal P}_{2}(ij)
+ \epsilon_4 {\cal P}_{4}(ij)\right)
\end{equation}
which acts in the subspace of singly occupied sites.
Here,
${\cal P}_{F}(ij)$ projects neighboring sites at $ij$
into a state with total spin total spin $F$ and
$
\epsilon_0 = \frac{-4 J^2}{U_0+U_1 - 6 U_2}
$,
$
\epsilon_2 = \frac{-4 J^2}{U_0- 3 U_2}
$, and
$
\epsilon_4 = \frac{-4 J^2}{U_0+ 4 U_2}.
$
This effective hamiltonian is valid in the limit $J \ll U_0$.
We point out that ${\cal H}_{\rm eff}$ can also be written
in terms of powers up to four of spin operators on adjacent
sites ${\bf S}_i \cdot {\bf S}_j$, but we find the representation
in Eq.~(\ref{Eq:Heff}) easier to work with.

When all of the scattering lengths are equal, we
will have for the on-site parameters $U_1=U_2=0$, which will
make $\epsilon_0=\epsilon_2=\epsilon_4$.
This corresponds to the situation where the hamiltonian 
(\ref{Eq:Heff}) has a special
SU(5) symmetry.  For this situation,
the state with all sites having the same spin
wave function
$
\ket{\psi_{\rm MI}}=\prod_i A_{\alpha} \ket{\alpha}_i
$
will be an exact eigenstate of ${\cal H}_{\rm eff}$, which is
the ground state.  There will
still be a rather large degeneracy for this situation since
the ground state energy will not depend on the coefficients $A_{\alpha}$.
On the other hand, when the scattering lengths begin to
differ this degeneracy will be partially lifted.
We will now determine the phase diagram using the variational
wave function.  By evaluating
$\bra{\psi} {\cal H}_{\rm eff} \ket{\psi}$ and minimizing over
the set of five complex coefficients $A_{\alpha}$ we find
precisely the same coefficients as
in (\ref{Eq:variational}).  Moreover, the phase diagram
for this case will be identical with that given in Fig.~\ref{Fig:phased1}
with the the axes changed to
$a_{0,2}-a_{4}\rightarrow \epsilon_{0,2}-\epsilon_4$.
% End optical lattice part

At the level of virtual hopping Eq.~(\ref{Eq:Heff})
in perturbation theory (second order
in $J/U$), the states in phase $\cal N$
are degenerate with respect to the parameter
$\eta$.  However, this does not correspond to a true symmetry of the
parent hamiltonian.  One therefore expects
that this degeneracy will be removed at the ring exchange level
in perturbation theory (fourth order in $J/U$).  On the other hand,
quantum fluctuations at the virtual hopping level from
${\cal H}_{\rm eff}$ may favor particular phases which can be
of the same order as ring exchange.  The detailed treatment of
this problem is beyond the scope of the present work and will be
considered elsewhere \cite{turner06}.
% End optical lattice part

% Begin excitation discussion
Knowing the symmetry properties of the wave function leads to
a natural classification of the collective excitations of the system.
Breaking continuous symmetries leads to Goldstone bosons;
the number of such modes corresponds to the
number of broken symmetry generators.  Our hamiltonians will always
have an SO(3) symmetry which corresponds to global spin rotations
(at the point where all the scattering lengths are the same there
will be a special SU(5) symmetry).
As can be seen in the pictorial representation, 
the uniaxial nematic phase ${\cal N}1$
has one remaining continuous
symmetry and thus
break two of the generators of SO(3).  This state will therefore have two
Goldstone bosons.
The biaxial nematic phases ${\cal N}2$ and ${\cal N}3$,
and tetrahedratic phase $\cal T$ all have no remaining continuous
symmetries,
so these will have three Goldstone
modes.
The topological excitations out of the various phases
can also be classified using the
homotopy theory.  The fundamental group describes 
how different
defects combine.  For instance, it is known that 
the fundamental
group of the biaxial nematic ${\cal N}3$ is the (nonabelian)
quaternion group \cite{mermin79}. 
The phase ${\cal N}2$ (transforming as the square)
has the larger 16-element dicyclic group \cite{coxeter80} 
for its fundamental group, giving it more types of
topological excitations. 
Finally, the fundamental group of the tetrahedratic state is the binary 
tetrahedral group with 24 elements \cite{coxeter80}.  
The number of topological excitations (4, 6, and 6 respectively)
is the number of nontrivial conjugacy classes in each of these groups.  
% End excitation discussion

% Begin spin 3 part
\begin{figure}
\includegraphics[width=3.5in]{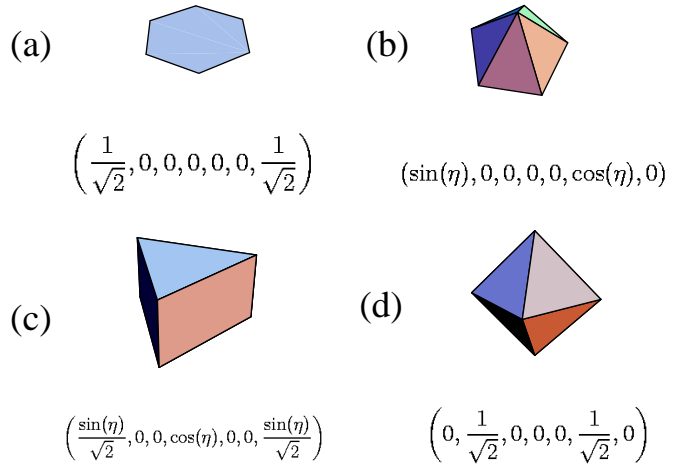}
\caption{Some possible  phases that can be realized for
a spin-three system either in the superfluid or Mott insulating states.
Shown are the coefficients of the wave functions $A_\alpha$ and the
shapes representing the symmetries of the wave functions.
The phases transform as the following polyhedra
(a) the hexagon, (b) the pyramid with pentagonal base,
(c) the prism,
and (d) the octahedron.  For phase (c) we have the condition
$\tan^2(\eta)<10$.  Additional phases similar to the spin two case
(not shown) are the ferromagnetic and nematic states.}
\label{Fig:spin3}
\end{figure}

We now will briefly comment on the types of states one can expect
for the spin-three
problem motivated by recent interest
\cite{griesmaier05,santos06,diener06}.
A similar variational wave function approach for either
the superfluid or Mott insulating states as done above can be used
for the spin-three case.
The characteristic polynomial $f_{\psi}(\zeta)$
will have six roots (instead of four as in the spin-two case).
Possible phases that can be realized for such a system are shown
in Fig.~\ref{Fig:spin3}.  The boundaries between these phases
can be found in \cite{santos06,diener06}; the purpose
here is to show that our classification scheme can be applied naturally
to the spin-three system.

%End spin 3 part

%Begin conclusion
In conclusion, we have presented a classification scheme for cold
spinor bosons.  We considered 
spin-two bosons as an illustrative example, which can be realized as
a hyperfine state of an alkali atom.  In addition to the
ferromagnetic and nematic states, we showed that a tetrahedratic
ordering can occur for this system.  Finally, we discussed an extension
to spin-three bosons.
%End conclusion

%Begin Acknowledgements
This work was supported by the NSF grant DMR-0132874 and
the Harvard-MIT CUA.
We thank T.-L. Ho, A. Imambekov, A. Kolezhuk, M. Lewenstein, and S. Sachdev
for useful discussions.
%End Acknowledgements.


\begin{thebibliography}{10}
\providecommand*{\bibinfo}[2]{#2}
\providecommand*{\eprint}[1]{#1}
\providecommand*{\url}[1]{#1}
\bibitem{lewenstein06}
\bibinfo{author}{M.~Lewenstein}, \bibinfo{author}{A.~Sanpera},
  \bibinfo{author}{V.~Ahufinger}, \bibinfo{author}{B.~Damski},
  \bibinfo{author}{A.~S. De}, and \bibinfo{author}{U.~Sen}, cond-mat/0606771.
\bibitem{stenger98}
\bibinfo{author}{J.~Stenger}, \bibinfo{author}{S.~Inouye},
  \bibinfo{author}{D.~M. Stamper-Kurn}, \bibinfo{author}{H.~J. Miesner},
  \bibinfo{author}{A.~P. Chikkatur}, and \bibinfo{author}{W.~Ketterle},
  \bibinfo{journal}{Nature} \bibinfo{volume}{\textbf{396}},
  \bibinfo{pages}{345} (\bibinfo{date}{1998}).
\bibitem{chang04}
\bibinfo{author}{M.~S. Chang}, \bibinfo{author}{C.~D. Hamley},
  \bibinfo{author}{M.~D. Barrett}, \bibinfo{author}{J.~A. Sauer},
  \bibinfo{author}{K.~M. Fortier}, \bibinfo{author}{W.~Zhang},
  \bibinfo{author}{L.~You}, and \bibinfo{author}{M.~S. Chapman},
  \bibinfo{journal}{Phys. Rev. Lett.} \bibinfo{volume}{\textbf{92}},
  \bibinfo{pages}{140403} (\bibinfo{date}{2004}).
\bibitem{chang05}
\bibinfo{author}{M.~S. Chang}, \bibinfo{author}{Q.~S. Qin},
  \bibinfo{author}{W.~X. Zhang}, \bibinfo{author}{L.~You}, and
  \bibinfo{author}{M.~S. Chapman}, \bibinfo{journal}{Nature Phys.}
  \bibinfo{volume}{\textbf{1}}, \bibinfo{pages}{111} (\bibinfo{date}{2005}).
\bibitem{schmaljohann04}
\bibinfo{author}{H.~Schmaljohann}, \bibinfo{author}{M.~Erhard},
  \bibinfo{author}{J.~Kronjager}, \bibinfo{author}{M.~Kottke},
  \bibinfo{author}{S.~van Staa}, \bibinfo{author}{L.~Cacciapuoti},
  \bibinfo{author}{J.~J. Arlt}, \bibinfo{author}{K.~Bongs}, and
  \bibinfo{author}{K.~Sengstock}, \bibinfo{journal}{Phys. Rev. Lett.}
  \bibinfo{volume}{\textbf{92}}, \bibinfo{pages}{040402}
  (\bibinfo{date}{2004}).
\bibitem{ho98}
\bibinfo{author}{T.~L. Ho}, \bibinfo{journal}{Phys. Rev. Lett.}
  \bibinfo{volume}{\textbf{81}}, \bibinfo{pages}{742} (\bibinfo{date}{1998}).
\bibitem{ohmi98}
\bibinfo{author}{T.~Ohmi} and \bibinfo{author}{K.~Machida},
  \bibinfo{journal}{J. Phys. Soc. Japan} \bibinfo{volume}{\textbf{67}},
  \bibinfo{pages}{1822} (\bibinfo{date}{1998}).
\bibitem{zhou04}
\bibinfo{author}{F.~Zhou}, \bibinfo{author}{M.~Snoek},
  \bibinfo{author}{J.~Wiemer}, and \bibinfo{author}{I.~Affleck},
  \bibinfo{journal}{Phys. Rev. B} \bibinfo{volume}{\textbf{70}},
  \bibinfo{pages}{184434} (\bibinfo{date}{2004}).
\bibitem{imambekov03}
\bibinfo{author}{A.~Imambekov}, \bibinfo{author}{M.~Lukin}, and
  \bibinfo{author}{E.~Demler}, \bibinfo{journal}{Phys. Rev. A}
  \bibinfo{volume}{\textbf{68}}, \bibinfo{pages}{063602}
  (\bibinfo{date}{2003}).
\bibitem{garcia-ripoll04}
\bibinfo{author}{J.~J. Garcia-Ripoll}, \bibinfo{author}{M.~A. Martin-Delgado},
  and \bibinfo{author}{J.~I. Cirac}, \bibinfo{journal}{Phys. Rev. Lett.}
  \bibinfo{volume}{\textbf{93}}, \bibinfo{pages}{250405}
  (\bibinfo{date}{2004}).
\bibitem{zhang04}
\bibinfo{author}{W.~X. Zhang}, \bibinfo{author}{S.~Yi}, and
  \bibinfo{author}{L.~You}, \bibinfo{journal}{Phys. Rev. A}
  \bibinfo{volume}{\textbf{70}}, \bibinfo{pages}{043611}
  (\bibinfo{date}{2004}).
\bibitem{bernier05}
\bibinfo{author}{J.-S. Bernier}, \bibinfo{author}{K.~Sengupta}, and
  \bibinfo{author}{Y.~B. Kim}, cond-mat/0510290.
\bibitem{widera05}
\bibinfo{author}{A.~Widera}, \bibinfo{author}{F.~Gerbier},
  \bibinfo{author}{S.~Folling}, \bibinfo{author}{T.~Gericke},
  \bibinfo{author}{O.~Mandel}, and \bibinfo{author}{I.~Bloch},
  \bibinfo{journal}{Phys. Rev. Lett.} \bibinfo{volume}{\textbf{95}},
  \bibinfo{pages}{190405} (\bibinfo{date}{2005}).
\bibitem{gerbier06}
\bibinfo{author}{F.~Gerbier}, \bibinfo{author}{A.~Widera},
  \bibinfo{author}{S.~Folling}, \bibinfo{author}{O.~Mandel}, and
  \bibinfo{author}{I.~Bloch}, \bibinfo{journal}{Phys. Rev. A}
  \bibinfo{volume}{\textbf{73}}, \bibinfo{pages}{041602}
  (\bibinfo{date}{2006}).
\bibitem{ciobanu00}
\bibinfo{author}{C.~V. Ciobanu}, \bibinfo{author}{S.~K. Yip}, and
  \bibinfo{author}{T.~L. Ho}, \bibinfo{journal}{Phys. Rev. A}
  \bibinfo{volume}{\textbf{61}}, \bibinfo{pages}{033607}
  (\bibinfo{date}{2000}).
\bibitem{ueda02}
\bibinfo{author}{M.~Ueda} and \bibinfo{author}{M.~Koashi},
  \bibinfo{journal}{Phys. Rev. A} \bibinfo{volume}{\textbf{65}},
  \bibinfo{pages}{063602} (\bibinfo{date}{2002}).
\bibitem{zawitkowski06}
\bibinfo{author}{L.~Zawitkowski}, \bibinfo{author}{K.~Eckert},
  \bibinfo{author}{A.~Sanpera}, and \bibinfo{author}{M.~Lewenstein},
  cond-mat/0603273.
\bibitem{zhou06}
\bibinfo{author}{F.~Zhou} and \bibinfo{author}{G.~Semenoff},
  \bibinfo{journal}{Phys. Rev. Lett.} \bibinfo{volume}{\textbf{97}},
  \bibinfo{pages}{180411} (\bibinfo{date}{2006}).
\bibitem{griesmaier05}
\bibinfo{author}{A.~Griesmaier}, \bibinfo{author}{J.~Werner},
  \bibinfo{author}{S.~Hensler}, \bibinfo{author}{J.~Stuhler}, and
  \bibinfo{author}{T.~Pfau}, \bibinfo{journal}{Phys. Rev. Lett.}
  \bibinfo{volume}{\textbf{94}}, \bibinfo{pages}{160401}
  (\bibinfo{date}{2005}).
\bibitem{santos06}
\bibinfo{author}{L.~Santos} and \bibinfo{author}{T.~Pfau},
  \bibinfo{journal}{Phys. Rev. Lett.} \bibinfo{volume}{\textbf{96}},
  \bibinfo{pages}{190404} (\bibinfo{date}{2006}).
\bibitem{diener06}
\bibinfo{author}{R.~Diener} and \bibinfo{author}{T.-L. Ho},
  \bibinfo{journal}{Phys. Rev. Lett.} \bibinfo{volume}{\textbf{96}},
  \bibinfo{pages}{190405} (\bibinfo{date}{2006}).
\bibitem{degennes95}
\bibinfo{author}{P.~B. de~Gennes} and \bibinfo{author}{J.~Prost},
  \bibinfo{title}{\emph{The Physics of Liquid Crystals}}
  (\bibinfo{publisher}{Oxford University Press}, \bibinfo{year}{1995}).
\bibitem{mermin79}
\bibinfo{author}{N.~D. Mermin}, \bibinfo{journal}{Rev. Modern Phys.}
  \bibinfo{volume}{\textbf{51}}, \bibinfo{pages}{591} (\bibinfo{date}{1979}).
\bibitem{bacry74}
\bibinfo{author}{H.~Bacry}, \bibinfo{journal}{J. Math. Phys.}
  \bibinfo{volume}{\textbf{15}}, \bibinfo{pages}{1686} (\bibinfo{date}{1974}).
\bibitem{klein03}
\bibinfo{author}{F.~Klein}, \bibinfo{title}{\emph{Lectures on the Icosahedron}}
  (\bibinfo{publisher}{Dover Publications}, \bibinfo{year}{2003}).
\bibitem{turner06}
\bibinfo{author}{A.~Turner}, \bibinfo{author}{R.~Barnett}, and
  \bibinfo{author}{E.~Demler}, to be published.
\bibitem{coxeter80}
\bibinfo{author}{H.~S.~M. Coxeter} and \bibinfo{author}{W.~O.~J. Moser},
  \bibinfo{title}{\emph{Generators and Relations for Discrete Groups, $4th$
  Ed.}} (\bibinfo{publisher}{Springer-Verlag}, \bibinfo{year}{1980}).
\end{thebibliography}
\end{document}